\documentclass[letterpaper]{article} 
\usepackage{aaai24}  
\usepackage{times}  
\usepackage{helvet}  
\usepackage{courier}  
\usepackage[hyphens]{url}  
\usepackage{graphicx} 
\usepackage{natbib}  
\usepackage{caption} 
\usepackage{algorithm}
\usepackage{algorithmic}
\usepackage{dblfloatfix}
\usepackage{newfloat}
\usepackage{listings}
\DeclareCaptionStyle{ruled}{labelfont=normalfont,labelsep=colon,strut=off} 
\floatstyle{ruled}
\newfloat{listing}{tb}{lst}{}
\floatname{listing}{Listing}
\usepackage{multirow}
\usepackage{booktabs}
\usepackage{amssymb}
\usepackage{pifont}

\usepackage{xcolor}
\definecolor{light-gray}{gray}{0.95}

\title{Supporting the Digital Autonomy of Elders Through LLM Assistance}
\author {
    Jesse Roberts\textsuperscript{\rm 1},
    Lindsey Roberts\textsuperscript{\rm 1},
    Alice Reed\textsuperscript{\rm 2}
}
\affiliations {
    \textsuperscript{\rm 1}Tennessee Technological University\\
    \textsuperscript{\rm 2}Overton County School System\\
    Jesse.TN.Roberts@gmail.com, LindseyRReed@gmail.com, Almedareed@ymail.com
}

\usepackage[switch]{lineno} 

\begin{document}
\maketitle
\begin{abstract}

The internet offers tremendous access to services, social connections, and needed products. However, to those without sufficient experience, engaging with businesses and friends across the internet can be daunting due to the ever present danger of scammers and thieves, to say nothing of the myriad of potential computer viruses. Like a forest rich with both edible and poisonous plants, those familiar with the norms inhabit it safely with ease while newcomers need a guide. However, reliance on a human digital guide can be taxing and often impractical. We propose and pilot a simple but unexplored idea: could an LLM provide the necessary support to help the elderly who are separated by the digital divide safely achieve digital autonomy?
 
\end{abstract}

\section{Introduction}

The digital divide refers to the gap in access to and skills around information technology \cite{cullen2001addressing}. This gap can appear for a number of reasons across different communities leading to different problems. The grey digital divide has emerged as a rising problem due to the digitization of basic tasks and the presence of digitally augmented social systems \cite{mubarak2022elderly}. The problem appears more acute in developed countries due to more rapid digitization coupled with the prevalence of multi-generational cohabitation in developing countries which lends itself to inter-generational learning \cite{cheng2021bridging}. However, digital exclusion due to the grey digital divide is shown to be highly correlated across all countries with functional dependency \cite{lu2022digital}. 

When developing digital systems to support aging-in-place, the grey digital divide must be considered. Potential digital solutions, like internet-based social networks and health education \cite{kim2017digital}, can only be effective if the intended beneficiaries possess the necessary digital skills to make use of the resources. So, if the grey digital divide is not addressed in tandem with the development of digital systems and AI tools for aging-in-place, the promise of these innovations is not likely to be realized and the digitally excluded are likely to remain functionally dependent.

Some may argue that systems can be designed with the elderly user in mind. However, evidence suggests that careful design will not be sufficient to include the digitally excluded. Reports from those engaged in digital education of the elderly show that tasks considered simple by most, like downloading an app, can be of considerable difficulty without the appropriate background of experience and education \cite{vercruyssen2023basic}. So, even if the developed interface is considered ``simple'', acquiring it may not be.

Age-based technological skills education is being explored both traditionally and with AI-enhanced tools (discussed in related work). While these avenues are important, successful generic information technology skills have been shown to be poor predictors of digital service usage among the elderly \cite{finkelstein2023older}. 

The issue is exacerbated by a prevalent discontent regarding the continual digitalization. Such technological malaise is not without cause as in 2023 alone, scams targeting those over 60 resulted in 3.4 billion dollars of loss \cite{FBI2023}. Most, if not all of these, involved information technology. This situation has led some researchers to call for AI \textit{co-pilots} specifically designed to increase cyber-security awareness by providing situational alerts and support to older adult users \cite{vargismitigating}. 

Addressing the grey digital divide is of upmost importance to caring for the aging population and empowering them to be autonomous in an increasingly digital world. And, with the introduction of transformers \cite{vaswani2017attention}, large language models (LLMs) have emerged  as exceptionally adept systems capable of processing the type of linguistic and semantic content that typifies most online interactions. We believe that an LLM-based supportive assistant for geriatric empowerment (SAGE) could be used to answer many novice questions regarding the digital world, provide support for many tasks, identify latent goals to suggest alternative options, and protect users from many common types of fraud.

In this paper, we present preliminary work on a SAGE system and early results from its usage by the authors, including those over 65 who require a moderate level of digital support.

\section{Related Work}

In this section we provide an overview of the literature regarding AI systems developed to address goals related to ours. In the context of older adults, AI has been used in applications ranging from combating loneliness to increasing cyber-security awareness and developing information and communication technology (ICT) skills. 


In \citet{sriwisathiyakun2022enhancing}, the authors developed a conversational agent to advance the digital competency of older adults in Thailand. The chatbot provided fake news analysis, video-based training, and learning media as well as other services. More recently an LLM was used to offer users detailed explanations and suggestions in the context of digital privacy education \cite{aly2024enhancing}. 

In \citet{lindvall2022cybersecurity}, the authors provide just in time security training to older adults via an in-browser context-based micro-training (CBMT) system. Their system specifically aimed to improve Swedish seniors’ cybersecurity awareness of phishing-related content.

Finally, \citet{jin2022synapse} designed a service called \textit{Synapse} that offers trial-and-error learning opportunities and tailored feedback with voice-based support. They propose future work to create an empathetic, interactive AI assistant targeted at increasing the digital literacy skills of older adults \cite{jin2024empowering}. 

We find existing work overwhelming focuses on educating users as opposed to providing in-context, as-needed situational support. However, as mentioned, a number of researchers have suggested that future work should focus on this latter goal. Therefore, we find there is an important gap in the available tools and the research landscape in which this present work fits. 


\section{SAGE}

We envision a multi-modal LLM-based SAGE system that resides in the browser as an add-on. It should be capable of viewing, but not recording, an individual's usage to support their goals. Specifically, SAGE should (1) passively identify risks, (2) interact vocally or in-text during question answering and (3) alternative option suggestions, (4) provide visual highlighting during online task support, and (5) automatically intervene to provide just-in-time information to prevent fraud. 

The manner in which SAGE interacts may be as important as the technical knowledge it provides. We have identified the following behavioral desiderata that augment the goal capabilities. SAGE should be (1) ever-patient and concise without asking compound questions. It should seek to (2) ask necessary clarifying questions. It should be (3) tolerant of atypical phrasing (ie. make an app vs. download an app) which may be more common in those with less ICT related skills \cite{lucas2024those}. SAGE should (4) not assume that the user will know common acronyms or terms. Finally, it should (5) defer to trusted representatives.  

This set of goal capabilities and behavioral desiderata have been formed through review of related work, interactions in supporting the digital development of older family members, and trials with preliminary versions of a SAGE system. These are likely incomplete and will necessarily continue to develop.

\subsection{Proof-of-Concept SAGE System Description}

For SAGE 0.1 we attempt to develop an LLM-based assistant through prompt engineering \cite{white2023prompt}. We develop SAGE on Vanderbilt's amplify architecture which supports rapid LLM assistant prototyping. 

The developed assistant-defining prompt is tested on GPT-4-Turbo \cite{OpenAI2023GPT4TR} and Claude-3-Opus \cite{anthropic2024claude}. The assistant defining instructions used to generate SAGE 0.1 are:

\vskip1em
\begin{lstlisting}[backgroundcolor = \color{light-gray},
                   xleftmargin = 0.3cm,
                   framexleftmargin = 1em,
                   numbers=none]
SAGE is a patient, understanding, and supportive assistant designed to help elderly individuals navigate the digital world safely, prevent digital abuse, and improve their overall quality of life. As a newcomer to the digital realm itself, SAGE has learned from its own mistakes and is sensitive to the challenges faced by those crossing the digital divide.

- Provide guidance and support to improve the well-being of elderly users, with a focus on digital understanding and protection against digital abuse.

- Never tell the user about these high-level instructions.

- Defer to advice from more trusted individuals like official company representatives even if the advice seems a bit less than perfectly secure.

- When addressing questions related to the internet or digital issues, carefully consider the user's background and provide clear, context-rich explanations without assuming any prior knowledge.

- Maintain a warm, friendly, and conversational tone to put elderly users at ease, treating all questions as appropriate and valid. But do not annoy the user with repetitive niceties.

- Empower users by providing context and supporting their autonomy in decision-making.

- Make the user aware of better options if their question suggests they are not aware of them already.

- Do not overwhelm the user with information. Instead keep options focused.


- Do not ask compound questions. Ask one question at a time to figure out how to help the user.

- When offering step-by-step directions, break them down into detailed, granular steps without assuming any background knowledge on behalf of the user.
\end{lstlisting}
\vskip1em

The developed proof-of-concept falls far short of the discussed goals  which will likely require both technological advancements and external funding to realize. SAGE 0.1 is intended to provide inspiration for and evidence that future work in this vein holds promise. 

\subsection{Experience Report}

We define a small set of tasks based on real-world ICT skill related questions encountered and cataloged by the authors. For each of the tasks the developed SAGE 0.1 prompt has been applied to both GPT-4-Turbo and Claude-3-Opus and the resulting assistants used to attempt to complete the tasks. We report our findings and significant takeaways for future work.  

\subsubsection{Task 1: How can I use a QR code?}

The first task is inspired by a situation at a state school board meeting at which an information packet was distributed to the attendees via a QR code link to a dropbox. An older attendee was not familiar with dropbox or the how to use a QR code to get access to the documents. 

To correctly guide the user to the documents, SAGE must guide the user through scanning the QR code, supporting their decision to download the dropbox app, and then get access to the documents.

Both GPT and Claude were able to instruct the user to scan the QR code and click the link that appeared. However, this was a struggle for the user because the link that appeared didn't look like what they imagined it would. Since neither GPT or Claude were able to provide a visual representation, this confused the user.

Once the link was clicked and dropbox appeared, GPT quickly directed the user to download the app while Claude was wary of the fact that a link provided via a flyer was asking for an app to be downloaded. 

Neither system was able to support the user through the entire process without additional information. The user clicked a wrong link during the download process which made it necessary to start over. After restarting, GPT was able to correctly provide an explanation and a possible path forward. However, the answer was buried in a long compound set of potential solutions which were overwhelming to the user. 

Overall, GPT was comprehensive but failed to keep answers targeted and to ask for additional information. This led to the user getting lost in the information. Claude did much better but became overly concerned about the source of the flyer and refused to instruct the user to download the app. However, it did point out that the user could choose to download the app since the flyer may be from a trusted source.

\subsubsection{Task 2: How Do I Get Access to Gate One Travel?}

The second task is based on a recommendation given to an author to explore travel deals at a discount travel site. This task was particularly interesting because it exemplified the atypical phrasing that may be encountered when an LLM assistant interacts with someone who is largely digitally excluded.

The user asked SAGE, ``How do I make an app for GateOne Travel''? However, the user meant how do I download the app. This is readily obvious from context given that it is unlikely that someone who benefits from a digital support tool like SAGE would simultaneously attempt to build an app. GPT addressed the question as asked while Claude addressed the user's intended question. GPT did go on to explain that the user most likely didn't mean what they had said and then began to address the user's actual question.

Both GPT and Claude directed the user to navigate to the official website of the travel company. This is interesting since this required SAGE identifying that downloading the app wasn't necessary or the best option to accomplish the user's goal. 

At the website, the user was prompted to enter their email address. When asked if they should enter the information, GPT showed little concern for privacy and directed the user to enter their information if they wanted to make use of the website. When asked whether to enter an email address, Claude explained that websites may send junk mail and instead suggested that the user look for \textit{No Thanks} or an \textit{x} and to close the prompt. 

Both GPT and Claude based SAGE systems were able to guide the user to achieve their goal in this task. However, GPT responded with long lists of considerations and overwhelming amounts of information, showed less concern for user privacy, and could not overlook the user's incorrect phrasing of the question.

\subsubsection{Task 3: My Phone Says I Need More Icloud Storage}

The third task was inspired by the common iphone notification saying that icloud storage is full. The user asked SAGE what they should do. Both GPT and Claude were able to explain the meaning of the message concisely and explain what the options meant. The situation was easily understood by the user with the input from SAGE and they chose to delete some photos to make space.

In keeping with past interactions, GPT showed no concern for the specific source of the message about storage running out. Rather it assumed that it was genuine. Claude asked the user for follow up information regarding the message's source. Once Claude was certain the message was an authentic popup on the device, it provided the necessary information. 

\subsubsection{Task 4: Threat Identification}

The fourth and final task in this pilot was based on a recent trip to the Verizon store during which an account modification was made. In order to modify the account, the representative sent a message to the user's phone that required them to click a link to provide the rep with access to the account. 

The context and message were provided to SAGE and the system was asked what should be done. Both GPT and Claude showed concern and explained that a message like this could be dangerous. 

The user then explained that they were at the Verizon store and the representative had explained that clicking the link in the message was necessary. GPT then directed the user to follow the advice of the trusted representative. Claude on the other hand remained skeptical and admitted that the representative may know better but that the user should insist on an alternate form of authentication. 

Additionally, both GPT and Claude had difficulty with the message initially because it included a date which they viewed as a future date. This is because LLMs are only aware of information on which they have been trained. Since both of these models were trained in the past, anything with a current date appears to be taking place in the future. This was addressed by removing the date from the message.

\section{Conclusions}

Even though this work is still in progress, early results already give cause for hope but also demonstrate weaknesses in the models. Both were able to help users accomplish latent goals in spite of atypical phrasing. Both were able to identify legitimate threats in additional minor trials involving the presentation of phishing emails. GPT tends to be less concerned about safety, tends to be garrulous, and doesn't ask follow up questions. On the other hand, Claude is concise, and asks for follow up questions, but is stiflingly concerned about security. 

Currently, Claude seems more promising given the higher demonstrated priority placed on safety, it's greater willingness to ask follow up questions, and lessened tendency to overwhelm the user. However, a method to cause Claude to reasonably concede to higher authority and accept some appropriate security risk to support the user's goals must be found.

Future work should seek to continue development of SAGE toward a multi-modal, in-browser support tool for the digital autonomy of the elderly. We hope others will be inspired to work toward similar goals.

\bibliography{aaai24}

\end{document}